\documentclass[aps,prb,twocolumn,amsmath,amssymb,superscriptaddress]{revtex4-1}

\usepackage{graphicx}
\usepackage{graphics}
\usepackage{dcolumn}
\usepackage{bm}
\usepackage[caption=false]{subfig} 
\usepackage{color}
\usepackage{tikz}
\usepackage{subfig}
\usepackage[titles,subfigure]{tocloft}

\newcommand{\beq}{\begin{eqnarray}}
\newcommand{\eeq}{\end{eqnarray}}
\newcommand{\dg}{\dagger}
\newcommand{\bpm}{\begin{pmatrix}}
\newcommand{\epm}{\end{pmatrix}}

\newcommand{\ket}[1]{| #1 \rangle}
\newcommand{\bra}[1]{\langle #1 |}
\newcommand{\dirac}[2]{\langle #1 | #2 \rangle}
\newcommand{\sub}{\subsection}

\usetikzlibrary{decorations}
\usetikzlibrary{decorations.pathreplacing}
\usetikzlibrary{decorations.markings}

\definecolor{purple}{rgb}{0.5,0,0.5}
\definecolor{dkgreen}{rgb}{0,0.5,0}

\begin{document}

\title{`Holographic' treatment of surface disorder on a topological insulator}

\author{Kun Woo Kim}
\affiliation{School of Physics, Korea Institute for Advanced Study, Seoul 130-722, Korea}
\author{Roger S. K. Mong}
\affiliation{Department of Physics and Institute of Quantum Information and Matter, California Institute of Technology, Pasadena, CA 91125, USA}
\affiliation{Department of Physics and Astronomy, University of Pittsburgh, Pittsburgh, PA 15260, USA}
\author{Marcel Franz}
\affiliation{Department of Physics and Astronomy, University of British Columbia, Vancouver, British Columbia, V6T 1Z1, Canada}
\author{Gil Refael}
\affiliation{Department of Physics and Institute of Quantum Information and Matter, California Institute of Technology, Pasadena, CA 91125, USA}

\date{\today}
\begin{abstract}
The effect of surface disorder on electronic systems is particularly
interesting for topological phases with surface and edge states. Using
exact diagonalization, it has been demonstrated that the
surface states of a 3D topological insulator survive strong surface
disorder, and simply get pushed to a clean part of the bulk. Here we
explore a new method which analytically eliminates the clean bulk, and
reduces a $D$-dimensional problem to a Hamiltonian-diagonalization
problem within the $(D-1)$-dimensional disordered surface. This dramatic
reduction in complexity allows the analysis of significantly bigger
systems than is possible with exact diagonalization. We use our method to analyze a 2D
topological spin-Hall insulator with non-magnetic and magnetic edge
impurities, and we calculate the probability density (or local density
of states) of the zero-energy eigenstates as a function of edge-parallel momentum and layer index.
Our analysis reveals that the system
size needed to reach behavior in the thermodynamic limit increases with
disorder. We also compute the edge conductance as a function of disorder
strength, and  chart a lower bound for the length scale marking the
crossover to the thermodynamic limit. 
\end{abstract}
\maketitle

\section{Introduction}

Solid state systems inevitably contain impurities. Study of the impurity effects is of special importance in topological insulators (for a review see, e.g.\ Ref.~\onlinecite{hasa2010, hasa2011, fran2013, qi2011}), because their surface states are expected to be fundamentally robust with respect to certain types of disorder. Understanding how impurities affect these systems also has practical implications because 
the fabrication of topological insulators (TIs) requires fine tuning of the doping concentration to place the Fermi energy within the band gap\cite{hsie2009, chen2009,xia2009} or to induce superconductivity\cite{wray2011,sasa2011,hsie2012}. 
Impurities that are located at or near the TI surface are especially interesting. One relevant example is 
the treatment with $\mathrm{NO}_2$ (see supplement of Ref.~\onlinecite{hsie2009}) on the surface of as-grown topological insulator $\mathrm{Bi}_{2-x}\mathrm{Cu}_x\mathrm{Se}_3$ which is necessary to prevent the surface band bending caused by the adsorption of residual atoms present in the vacuum chamber. As the manifestations of topological systems -- such as the protected surface states or Majorana fermions -- are localized near the surface or the edge of the system \cite{mour2012,das2012, alic2012, been2013, elli2015}, the effect of surface impurities on the surface spectral and transport properties are of both theoretical and practical interest\cite{schu2012, pien2012, rain2013}.
Local density of states (LDOS) can be efficiently probed by scanning tunneling spectroscopy while the surface spectral function can be extracted from 
angle resolved photoemission spectroscopy (ARPES) measurements. Various techniques have been employed to study the transport properties of TI surfaces.

Recently, an exact diagonalization (ED) approach has been applied to non-interacting TIs with surface disorder\cite{schu2012,quei2014}. This study found evidence for a crossover between a nearly ballistic response of the surface electrons at weak disorder, localization physics at intermediate disorder, and then a restored nearly ballistic surface state hiding in the second layer when disorder is very strong. Further investigations of this phenomenon using ED are going to prove very challenging: the computational cost for analyzing the surface physics can be very high because the bulk of the system---although gapped---must be treated on equal footing with the surface degrees of freedom. The system sizes accessible to ED analysis are, therefore, rather small.
In particular, we demonstrate here that when strong surface impurities are present in the system, the lower bound of system size required to clearly resolve the bulk electronic properties and their effect on the surface states becomes large in proportion to the impurity potential strength (at least in two-dimensions). Treating sufficiently large systems using ED becomes computationally challenging in this limit and new techniques to address the problem are required. In this manuscript we develop such a technique. We also emphasize the need to treat sufficiently large systems in order to distinguish system properties 
at the thermodynamic limit versus those of the ``quantum dot'' regime, where finite size effects dominate.

In this manuscript, we introduce a new technique that allows us to efficiently extract the surface state properties of surface-disordered TIs. We obtain properties such as the surface spectral function, LDOS and transport properties of the surface channels, by essentially ``integrating out'' the clean bulk degrees of freedom analytically and obtain the effective surface-Hamiltonian describing a TI surface with arbitrarily strong impurities. This approach not only allows us to reduce the computational difficulty by one dimension (e.g.\ for a 3D TI with surface disorder we only need to solve a 2D problem), but also allow us to map a strong disordered problem into a weak disordered one where perturbation theory is valid. By constructing a self-consistent transfer-matrix approach, we are able to recover the exact energies and wavefunctions of the surface states both at the disordered layer and in the remaining bulk layers. 

The manuscript is organized as follows. In Sec.~\ref{sec:gene_fram}, we first explain how to integrate out the clean bulk degrees of freedom.
In Sec.~\ref{sec:appl_to} we then introduce a generic model Hamiltonian of a 2D TI with impurities on one of its edges. While our method is applicable to any layered system in arbitrary dimension, we choose to concentrate on the 2D case since it is both simple to present and analyze. Next, we report a series of results that clearly differentiate the 0D quantum-dot regime from the bulk regime (where finite-size errors are suppressed) and specify the lower bound on the system size to observe the latter for magnetic and non-magnetic edge impurities. We conclude the section with a discussion of the surface properties. 
Lastly, in Sec.~\ref{sec:tran_beha}, the conductance through the 2D TI edge channels is computed. The latter provides complementary information to the spectral properties discussed in Sec.~\ref{sec:gene_fram}. We conclude with a discussion in Sec.~\ref{sec:conc_and}.

\begin{figure}[h]
\begin{center}
\includegraphics[width=85mm]{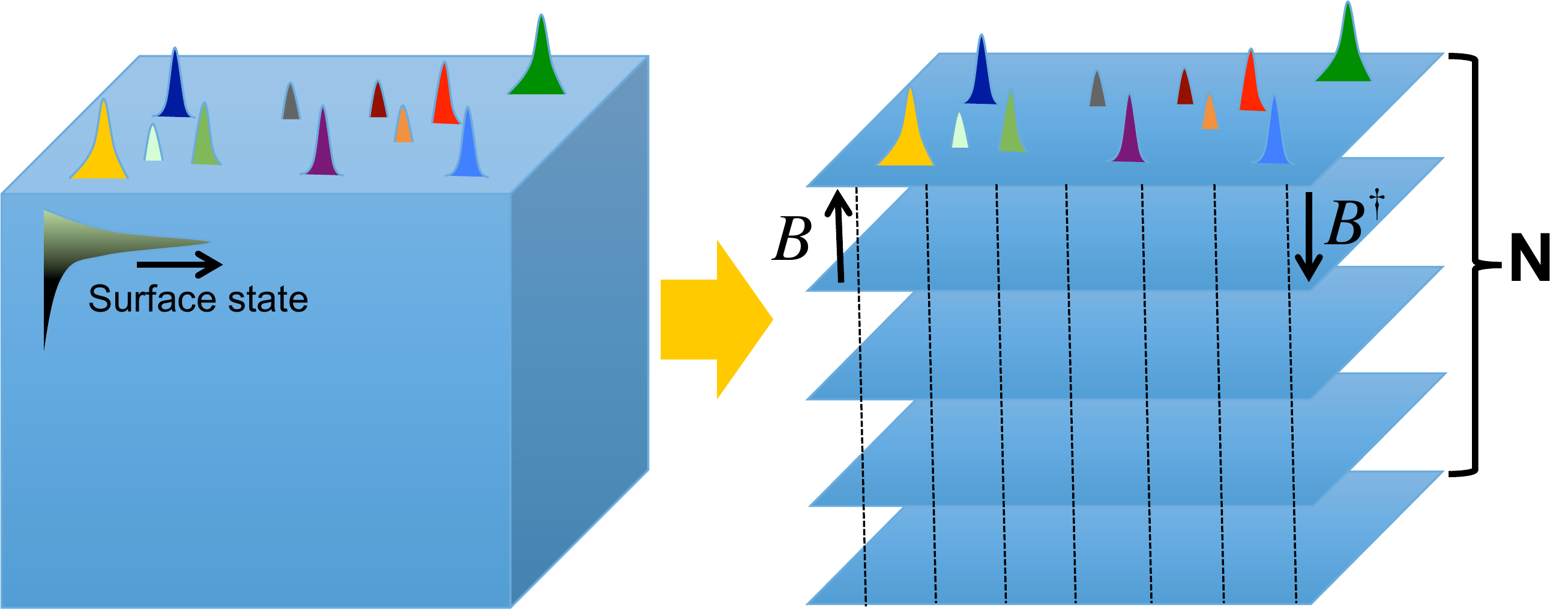}
\caption{A pictorial description of a system with impurities on the surface only. We are interested in studying the local density of surface state as the impurity strength increases. Our strategy is to decompose the system into clean layers, coupled through matrices $B$ and $B^{\dagger}$, and a surface containing impurities, and then analytically integrate out the former. }
\label{fig:3d_lattice}
\end{center}
\end{figure}

\section{General framework: effective single layer Hamiltonian}
\label{sec:gene_fram}

In this section, we provide a general derivation of our approach. Our goal is to exactly reduce diagonalization of a $D$-dimensional system with surface impurities to the diagonalization of an effective Hamiltonian describing just the $D-1$ dimensional surface. Taking the top surface to be disordered, we introduce a way to integrate out the clean layers from the bottom all the way to the top layer. This leaves us with a single layer effective Hamiltonian which includes the impurity potential and a self energy which accounts for the entire clean bulk.

\sub{Layered Schr\"odinger equation and self energy}
\label{sec:tran_matr}

We begin our analysis with the Schr\"odinger equation for layers parallel to the disordered surface
\beq
B\psi_{n-1} + [H_{0}+ V_\text{imp} \delta_{n,1}-E]\psi_{n} + B^\dagger \psi_{n+1}=0, \label{scho}
\eeq
where $\psi_{0}=0$, $\psi_n$ is a wavefunction on $n_{th}$ layer parallel to the impurity surface, $H_{||}$ is an in-layer Hamiltonian, $V_\text{imp}$ is an impurity potential in the first layer, $B$ and $B^\dg$ are hopping terms between layers, $n > 0$ is the layer index.
For notational convenience, we set $\psi_0=0$.
For $n=N$, the last layer of the system, we can write exactly
\beq
B\psi_{N-1} + [H_0-E] \psi_N = 0 \label{scho_last}
\eeq
Using the Schr\"odinger equation for the $n=N-1$ layer, $B\psi_{N-2} + [H_0-E]\psi_{N-1} + B^\dg \psi_N = 0$, and substituting Eq.~\eqref{scho_last}, we can``integrate" out the last ($N$'th) layer
\beq
B\psi_{N-2} + \left[H_0-E + B^\dg \frac{1}{E-H_0} B \right] \psi_{N-1}= 0. \label{scho_inte}
\eeq
Eliminating $\psi_N$ introduces for $\psi_{N-1}$ the effective potential $\Sigma_{N-1}=B^\dg \frac{1}{E-H_0} B $. By repeating this process, we can integrate out all layers up to the first layer and the following recursion relation can be found,
\beq
\Sigma_n = B^\dg\frac{1}{E-H_0-\Sigma_{n+1}} B \label{recu_back}
\eeq 
with a boundary condition $\Sigma_N = 0$. Recall that $B^\dg$ is a hopping to the next layer and $B$ is a hopping to the prior layer. And the effective potential $\Sigma_n$ is obtained by sandwiching the Green's function in $(n+1)_{th}$ layer by $B^\dg$ and $B$, describing a scattering process of hopping to the next layer, propagating, and hopping back to the original layer. 

Let us next write an effective Hamiltonian in the top layer in the following way:
\beq
[E- H_0 - \Sigma_1 ] \psi_1 = V_\text{imp} \psi_1 \\
\left[ B \Sigma_0 ^{-1} B^\dg \right] \psi_1 = V_\text{imp} \psi_1, \label{scho_S},
\eeq
where the recursion relation \eqref{recu_back} is again used to further simplify the clean part of the Hamiltonian. In the next section, we introduce a way to solve the recursion relation exactly.

\subsection{`Holographic' mapping of the self energy}

The recursion relation can be straightforwardly solved by mapping the effective potential to a matrix $M$ which obeys the same Schr\"odinger equation as the layer wavefunctions
\beq
BM_{n-1} + [H_{0}-E]M_{n} + B^\dagger M_{n+1}=0, \label{scho_M}
\eeq
where the matrix $M_n$ has the same dimension as the Hamiltonian $H_{||}$, and it is invertible by construction.

With Eq.~\eqref{scho_M}, the recursion relation for the self-energy is easily solved
\beq
\Sigma_n = B^\dg M_{n+1} M_{n}^{-1}. \label{sigm_M}
\eeq
One can directly verify that this is a solution of the recursion relation for $M_n$ satisfying boundary condition $M_{N+1}=0$. For a clean bottom-surface, we can exactly construct $M_n$ for a system with finite thickness (the calculation is detailed in the Appendix). 

The last step involves writing a close-form equation for the wavefunction of the top (disordered) layered. Note that as $M_n$ is also a solution of the Schr\"odinger equation, an element of $M_n$ scales with eigenvalues of transfer matrix of Schr\"odinger equation \eqref{scho_M}: $(M_n)_{ii'} \sim \rho_j^{N-n}$. Using the exact expression of $M_n$'s, we construct the left side of Eq.~\eqref{scho_S}. Then, we obtain the Schr\"odinger equation expressed in terms of $M_n$'s
\beq
\left[ B M_0 M_1^{-1} \right] \psi_1 = V_\text{imp} \psi_1. \label{Hami_redu}
\eeq
This is the effective single layer Hamiltonian. The left-hand side contains only elements from the clean part of the Hamiltonian, and involves the self energy from all subsequent layers; the right-hand side is simply the surface impurity potential operating on the top-layer wavefunction. $M_0$ and $M_1$ are a function of energy, and one can find all eigenvalues of a system by finding the energies that satisfy $\operatorname{det}\left[ B M_0 M_1^{-1} - V_\text{imp}\right] =0$. The surface wavefunction can be subsequently found from Eq.~\eqref{Hami_redu}, which is identical to the result from exact diagonalization. To obtain whole wavefunction in the layers beneath the top layer, we apply the transfer matrix which is also obtained in terms of $M_n$ as shown in the next section.

\subsection{Transfer matrix of wavefunctions}

The first layer wavefunction can be exactly obtained from Eq.~\eqref{Hami_redu}, therefore, the computational complexity is essentially reduced by one dimension. To obtain a full profile of the wavefunction in the subsequent layers, we construct an approach similar to the transfer matrix approach in this section. We will use the term ``transfer matrix'' quite liberally in what follows.

$\Sigma_n$ plays the role of effective potential in the $n_{th}$ layer, induced by integrating out the $(n+1)_{th}$ layer up to $N_{th}$ layer. In the Schr\"odinger equation \eqref{scho}, such a contribution is accounted by the third term on the left side. Therefore, we have the following equality:
\beq
B^\dg \psi_{n+1} = \Sigma_n \psi_n.
\eeq
One can explicitly show this relation by the elimination method introduced in Eq.~\eqref{scho_inte}. The transfer matrix is conveniently expressed in terms of the $M_n$'s using Eq.~\eqref{sigm_M}
\beq
\psi_{n+1} =\left[ M_{n+1} M_n^{-1} \right] \psi_n.
\eeq
Or, more generally, using the relation between the wavefunctions in the $m$ and $n$ layers
\beq
\psi_{n} =\left[ M_{n} M_m^{-1} \right] \psi_m,\label{tran},
\eeq
where the exact expression of $T_{n,m}\equiv M_{n} M_m^{-1}$ is known. Note that the expression for the transfer matrices is disorder-free, which implies that disordered wavefunction in the first layer propagates into the subsequent layers just as a clean wavefunction would. This, of course, makes sense since only the top layer contains impurities. A conventional transfer matrix constructed from the top surface, however, would always contain impurity potentials, and therefore the construction of the whole wavefunction would not be as straightforward.

\subsection{`Holographic' mapping of the impurity potential}

For completeness, we address another question of interest: what is the effective impurity potential experienced by an electronic state in the bulk due to the surface impurity. This question can be answered in the same formalism introduced in earlier sections. To compute the effective impurity potential, we integrate out the first $(n-1)$ layers. The recursion relation for effective potential is
\beq
V_{n+1} = B \frac{1}{E-H_0-V_{n}} B^\dg \label{recu_forw},
\eeq 
with boundary condition $V_1 = V_\text{imp}$. The `holographic' mapping helps us to analytically derive a scaling behavior of the effective potential
\beq
V_n = B \tilde{M}_{n-1} \tilde{M}_n ^{-1},
\eeq
where the $\tilde{M}_n$'s are similarly constructed to satisfy the Schr\"odinger equation for the individual layers, \eqref{scho_M}, and to be invertible. However, their boundary condition is different from the previous clean case. $\tilde{M}_n$ has to be constructed such that the following condition is satisfied:
\beq
V_\text{imp} = B \tilde{M}_{0} \tilde{M}_1 ^{-1} .
\eeq
Since $V_\text{imp}$ is a random matrix, it is nontrivial to determine $\tilde{M}_0$ in general. But, we know the object $\tilde{M}_n$ propagates just like a clean wavefunction. Therefore, it is possible to deduce the scaling of $(V_n)_{ij}$ with respect to layer index n;  which includes the contribution from surface impurities as well as clean layers from the top to $(n-1)^{th}$ layer.


\section{Application to a 2D topological insulator}
\label{sec:appl_to}

In the previous section, we introduced a general transfer-matrix framework for computing the full wavefunctions of layered systems with surface impurities. In this section, a 2D topological insulator model\cite{bern2006} is employed to explicitly show how the local density of states can be computed in a system with edge impurities.

\begin{figure}[t]
\begin{center}
\includegraphics[width=90mm]{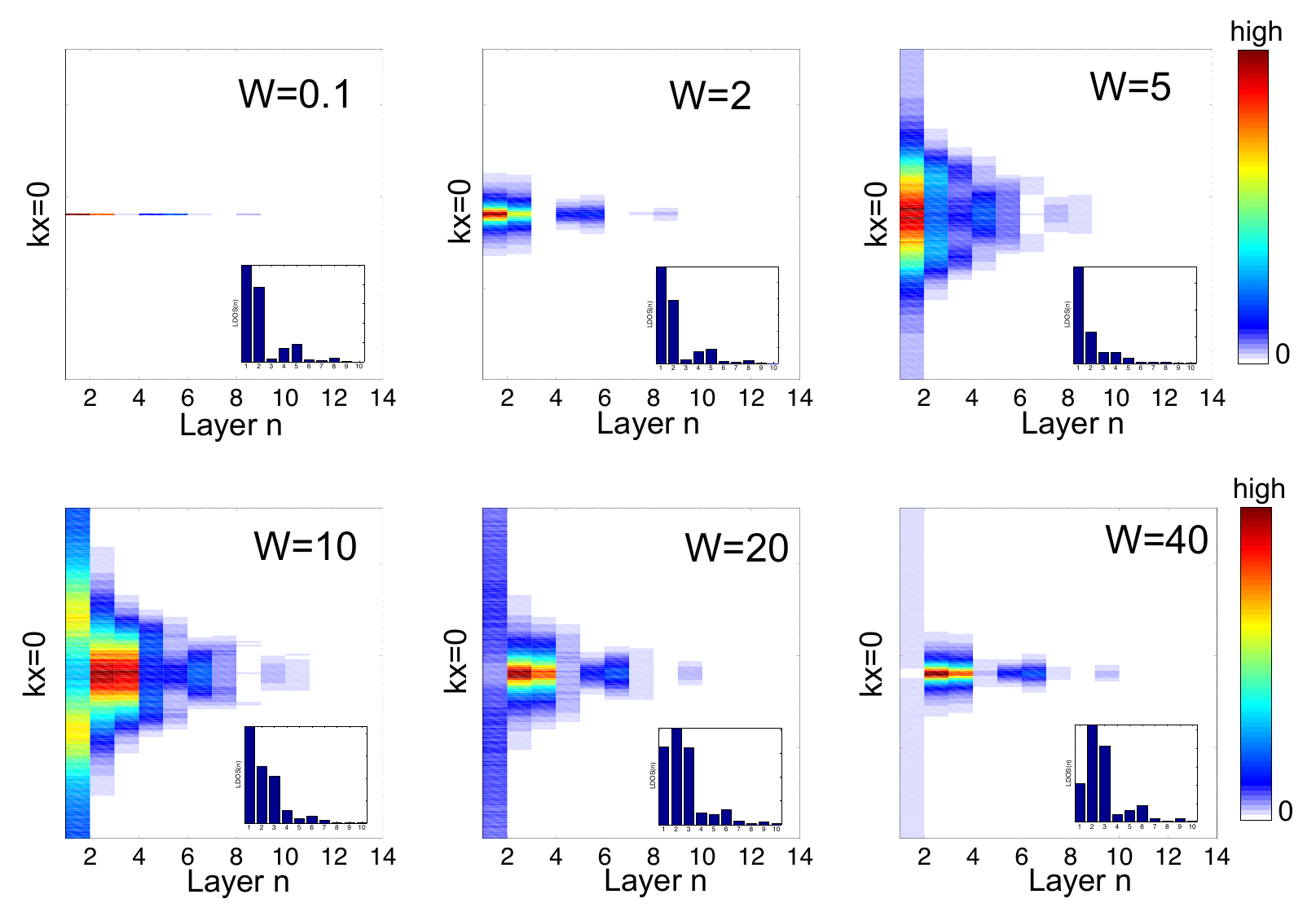}
\caption{The disorder-averaged LDOS at $E=0$ in wire index $n$ and momentum space $k_x$ of edge state at zero energy is plotted for increasing impurity strength. Starting from the clean edge state where the LDOS is only at $k_x=0$, the LDOS is spread out in momentum space then shifted to the second layer. The insets are showing the $k_x$-integrated LDOS as a function of layer $n$. Large enough system size $N_x=180$ is chosen such that the averaged LDOS is size-independent. 500 disorder realizations are averaged.}
\label{fig:vmap}
\end{center}
\end{figure}

\begin{figure}[t]
\begin{center}
\includegraphics[width=85mm]{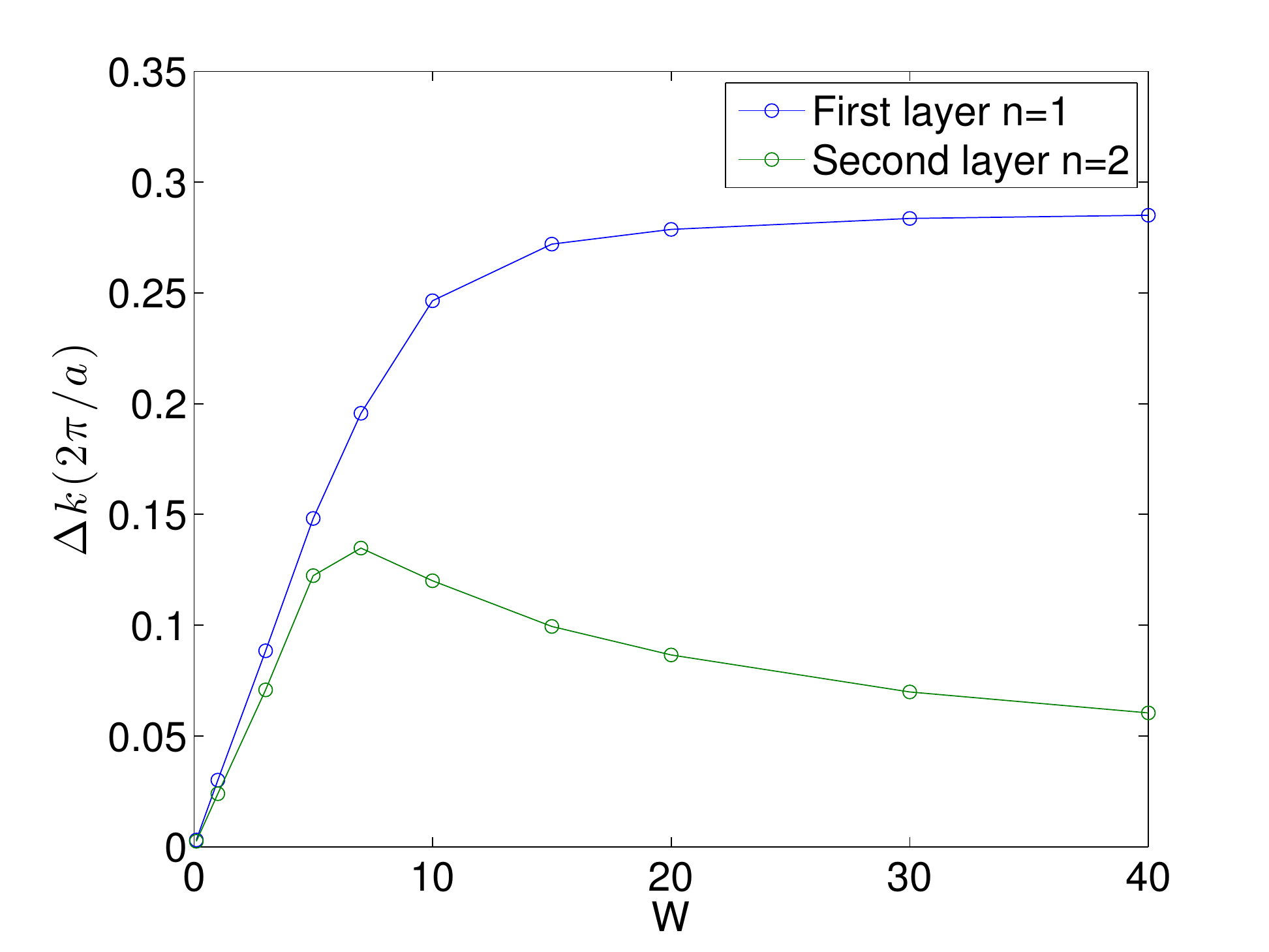}
\caption{
For system size $N_x=120$, the width of LDOS in the first and the second wire in momentum space $k_x$ is plotted for as a function of disorder strength $W$ for non-magnetic impurities. In clean system limit, the edge state LDOS $P(k_x)$ is concentrated at $k_x=0$ and its width is infinitesimally small. As the disorder strength increases, the LDOS spreads in momentum space. However, for the stronger disorder, the first layer LDOS takes on all possible momenta and its width saturates, while the second layer LDOS become concentrated around $k_x=0$ again.}
\label{fig:width}
\end{center}
\end{figure}

Our main results are presented in Fig.~\ref{fig:vmap}, \ref{fig:width} and \ref{fig:width_mag}, where we use the formalism developed earlier to compute the LDOS of the first and second layers of the TI varying the disorder strength $W$.

\sub{Model Hamiltonian}
\label{sec:mode_hami}

Consider the toy model of a 2D topological insulator. In momentum space,
\beq H(\vec{k}) = [m- 2b(2-\cos k_x-\cos k_y)]\tau_z \nonumber \\ + A[ \tau_x s_z \sin k_x+ \tau_y \sin k_y] \eeq
where $\tau_i$ is a Pauli matrix in orbital basis, $s_i$ a Pauli matrix in spin basis.
The lattice spacing is set to $a=1$ such that the momenta $k_x$ and $k_y$ lie within the interval $[-\pi,\pi]$.
To introduce an edge state, open boundary conditions are introduced in the $y$-direction, and periodic boundary conditions are applied to the $x$-direction.
The intra-layer Hamiltonian and the hopping term between layers described in Eq.~\eqref{scho} are:
\begin{align}\begin{split}
H_{0} &= [m- 2b(2-\cos k_x)]\tau_z +\tau_x s_z A\sin k_x, \\ B &= b \tau_z- i \frac{A}{2} \tau_y.
\end{split}\end{align}
The system is in the topological phase if the bands are inverted for some range of momentum: $\operatorname{sign}(mB)>0$. For this case, the dispersion of the top and bottom edge states are given by $E=\pm A\sin k_x$ \cite{imur2010}.

\begin{figure}[t]
\begin{center}
\includegraphics[width=75mm]{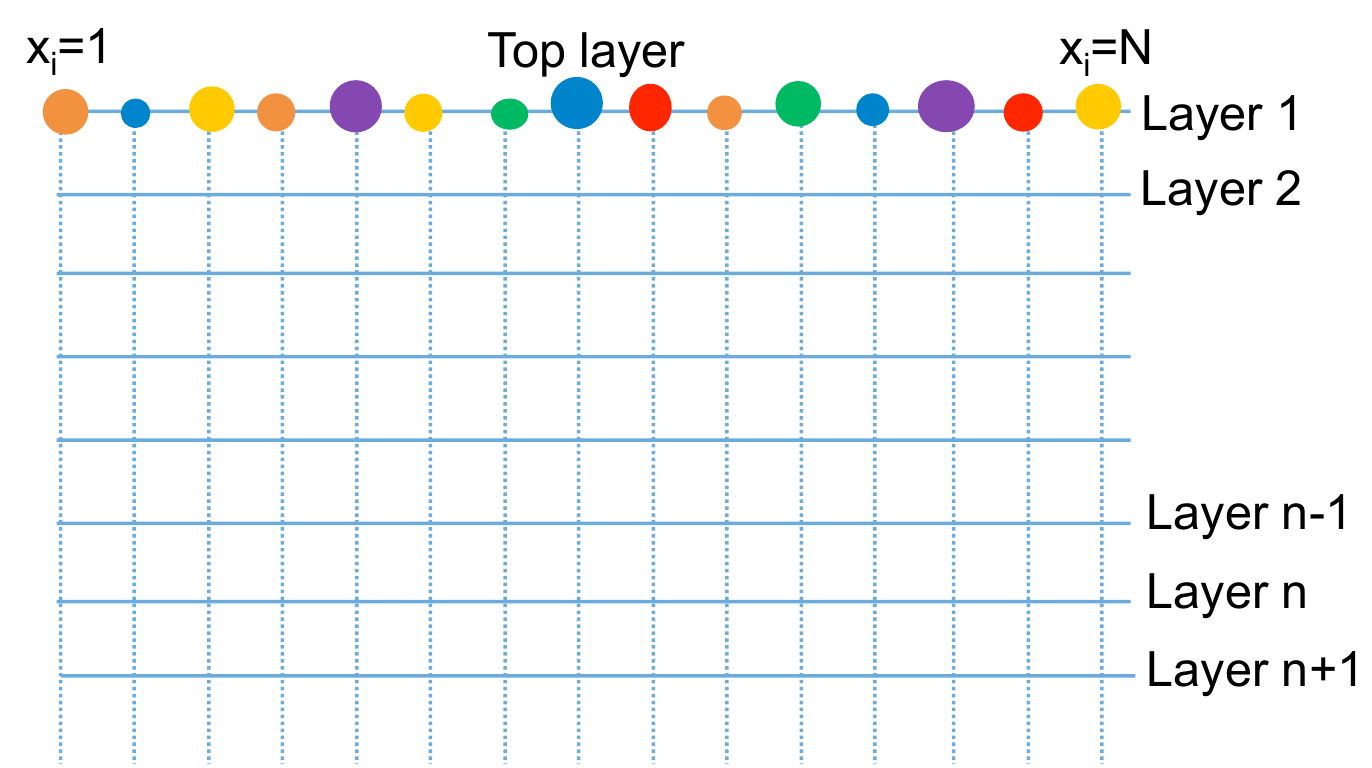}
\caption{A 2D lattice model for a TI with edge impurities. Along $x$-direction we have a periodic boundary condition that $x_1=x_{N+1}$, and along y-direction there are N-layers and the subscript $n$ indicates the $n_{th}$ layer along the $y$-direction. Different colors and sizes of dots on the top layer represent the random on-site impurity potentials. We study how the effect of the top edge impurities propagate down into the bulk layers. }
\label{fig:2d_lattice}
\end{center}
\end{figure}

\sub{A single layer effective Hamiltonian}\label{subsection:half_syst}

The system is equivalent to a set of parallel 1D wires coupled by a hopping matrix $B$ in spin and orbital basis.
We want to construct the matrix $M_n$ which is an essential building block of a single layer Hamiltonian [Eq.~\eqref{Hami_redu}] and the transfer matrix [Eq.~\eqref{tran}].
To demonstrate the method, we construct $M_n(k_x=0, E=0, s_z=1)$ here, and other $(k_x,E)$ cases will be shown in the appendix. Note that because the clean Hamiltonian is diagonal in momentum and spin space, we only need to analyze the orbital space. The Schr\"odinger equation we need to solve is
\beq
\left[ \tau_z \left( b\left( \frac{1}{\rho} +\rho\right) + m-2b \right) -i\tau_y
\frac{A}{2}\left( \frac{1}{\rho}-\rho\right) \right] \psi_n =0, \nonumber
\eeq
where we use $\psi_{n+m} = \rho^m\psi_n$. This is one section of Schr\"odinger equation at $k_x=0$, $E=0$, and $s_z=1$. There are four transfer eigenvalues and corresponding eigenvectors. By taking the determinant of the terms in the square bracket, we get $\rho_1=\lambda_1$, $\rho_2=1/\lambda_1$, $\rho_3=\lambda_2$, and $\rho_4=1/\lambda_2$, where
\beq
\lambda_1 &=& \frac{-(m-2b)- \sqrt{m^2-4mb+A^2}}{A+2b} \nonumber \\
\lambda_2 &=& \frac{-(m-2b) + \sqrt{m^2-4mb+A^2}}{A+2b} 
\nonumber
\eeq
where $\lambda_{1,2}$ is chosen to be $|\lambda_{1,2}|<1$ for $0<m<4b$ and corresponding eigenvectors are $\phi_1=\phi_3=\ket{+}$ and eigenvectors of $1/\lambda_{1,2}$ are $\phi_2=\phi_4=\ket{-}$, where
\beq
\ket{+} = \frac{1}{\sqrt{2}}\bpm 1 \\ 1 \epm, \ket{-} = \frac{1}{\sqrt{2}}\bpm 1 \\ -1 \epm .
\eeq
Considering only one spin section, this implies that a state $\ket{+}$ is localized at the top ($n=1$) and a state $\ket{-}$ is localized at the bottom ($n=N$).
The interlayer hopping operator can be expressed in terms of the eigenvectors: $B= (b-A/2)\ket{-}\bra{+} + (b+A/2)\ket{+}\bra{-}$. For this given set of eigenvalues and vectors, we can construct an invertible $M_n$ in the following manner:
\beq
M_{l+N+1}= (\rho_1^l - \rho_3^l) \ket{+}\bra{+} -(\rho_2^l - \rho_4^l)\ket{-}\bra{-} \nonumber ,
\eeq
where $l=n-N-1$. $M_{n}$ is invertible for $n\neq N+1$ and satisfies the homogeneous boundary condition $M_{N+1} = 0$. If $m^2-4mb+A^2<0$, eigenvalues are $\rho_1=\lambda e^{i\theta}$ and $\rho_3=\lambda e^{-i\theta}$ with $\lambda= |\frac{2b-A}{2b+A}|$. Thus, the effective Hamiltonian in the first layer is as follows\eqref{Hami_redu}: 
\begin{align}
&E-H^\text{eff}_1 = B M_0 M_1^{-1} \nonumber \\
&= B \left[ \frac{\rho_1^{-N-1}-\rho_3^{-N-1}}{\rho_1^{-N}-\rho_3^{-N}} \ket{+}\bra{+} + \frac{\rho_2^{-N-1}-\rho_4^{-N-1}}{\rho_2^{-N}-\rho_4^{-N}} \ket{-}\bra{-} \right] \nonumber \\
&= (b-A/2) \frac{1}{\lambda} \frac{\sin [(N+1)\theta]}{\sin (N\theta)} \ket{- }\bra{+} \notag \\
&\quad + (b+A/2) \lambda \frac{\sin [(N+1)\theta]}{\sin (N\theta)} \ket{+}\bra{-} ,
\label{Heff}
\end{align}
where $N$ is the number of parallel wires. The effective Hamiltonian contains off-diagonal elements only in $\ket{\pm}$ basis.

Note that the effective single-layer Hamiltonian at $(k_x,E)=(0,0)$ depends on the width of the system, and if a wavefunction contains a component at $(k_x,E)=(0,0)$, it will also be system-size dependent. Because we cannot think of a localized wavefunction dependent of the system size for large enough $N$, we can say no eigenstate localized to an edge sits at $(k_x,E)=(0,0)$. More relevant Hamiltonian sections at zero energy will be $k_x \neq 0$, which is system size independent in the large $N$ limit.
More generally, the Hamiltonian sectors not at $(k_x,E)=(2\pi l/N, A \sin k_x)$ is expressed in the following way: 
\begin{align}\begin{split}
\bra{+}H_{1}^\text{eff} \ket{+} &= \frac{b+A/2}{1/r_3-1/r_1} \left[ \rho^{-1}_1-\rho^{-1}_3 \right] , \\
\bra{+}H_{1}^\text{eff} \ket{-} &= \frac{b+A/2}{1/r_3-1/r_1} \left[ \rho^{-1}_1 /r_1 - \rho^{-1}_3/ r_3 \right],\\
\bra{-}H_{1}^\text{eff} \ket{+} &= \frac{b-A/2}{r_1-r_3} \left[ \rho_1^{-1} r_1 - \rho_3^{-1} r_3 \right] , \\
\bra{-}H_{1}^\text{eff} \ket{-} &= \frac{b-A/2}{r_1-r_3} \left[ \rho^{-1}_1- \rho^{-1}_3 \right]. 
\end{split}\end{align}
Here $\rho_{1,3}$'s are the eigenvalues of the transfer matrix with magnitude smaller than unity, and $r_i = \dirac{+}{\phi_i}/ \dirac{-}{\phi_i}$'s are the ratios of the overlaps between the transfer-matrix eigenstates corresponding to $\rho_i$ with the $\ket{+}$ and $\ket{-}$ states. We can see that as the $(k_x, E)$ approaches to the on-shell condition, $r_i$ approaches zero and the wavefunctions have infinitesimal overlap with $\ket{-}$ since $\bra{-}H_{1}^\text{eff} \ket{-} $ component is huge. In other words, the Hamiltonian expressed in this way can be interpreted as a projection to the on-shell eigenstates. 

\begin{figure}[b]
\begin{center}
\includegraphics[width=90mm]{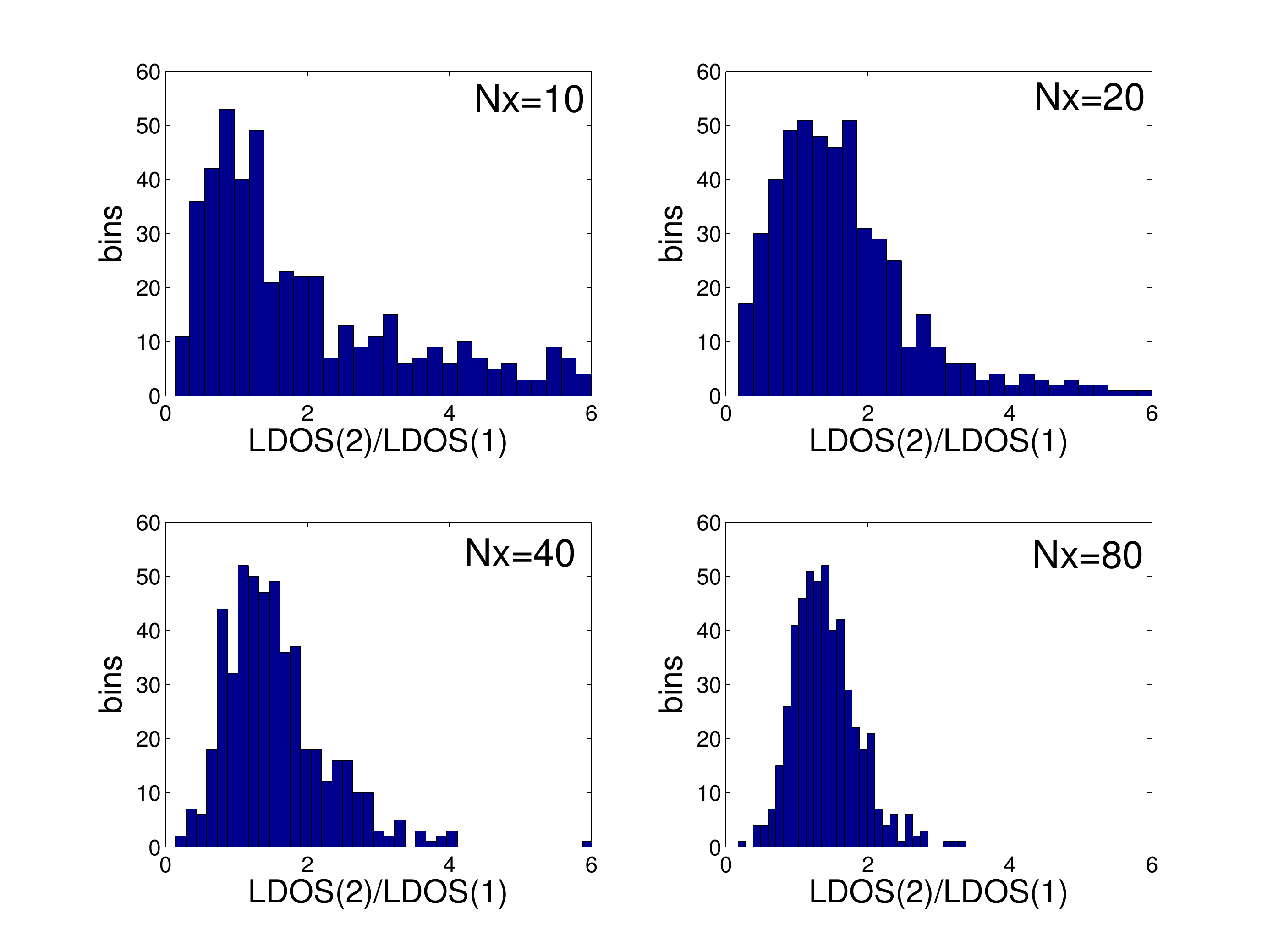}
\caption{For disorder strength $W=20$, histograms of the LDOS ratio for different system sizes $N_x=10,20,40,80$ are shown. When system size $N_x > 80$ the distribution converges and the ratio of the LDOS in the first and the second wire becomes system size independent, and the average of the LDOS ratio in the thermodynamic limit can be estimated. On the other hand, system sizes $N_x<40$ are in quantum-dot regime and not proper to compute the bulk electronic properties because of their system-size dependence.}
\label{fig:ldos_Nx}
\end{center}
\end{figure}

With a set of impurities on the top wire (Fig.~\ref{fig:2d_lattice}), to obtain eigenenergies we find energies where the determinant of effective single layer Hamiltonian is zero (see Eq.~\eqref{Hami_redu}).

\begin{figure}[t]
\includegraphics[width=90mm]{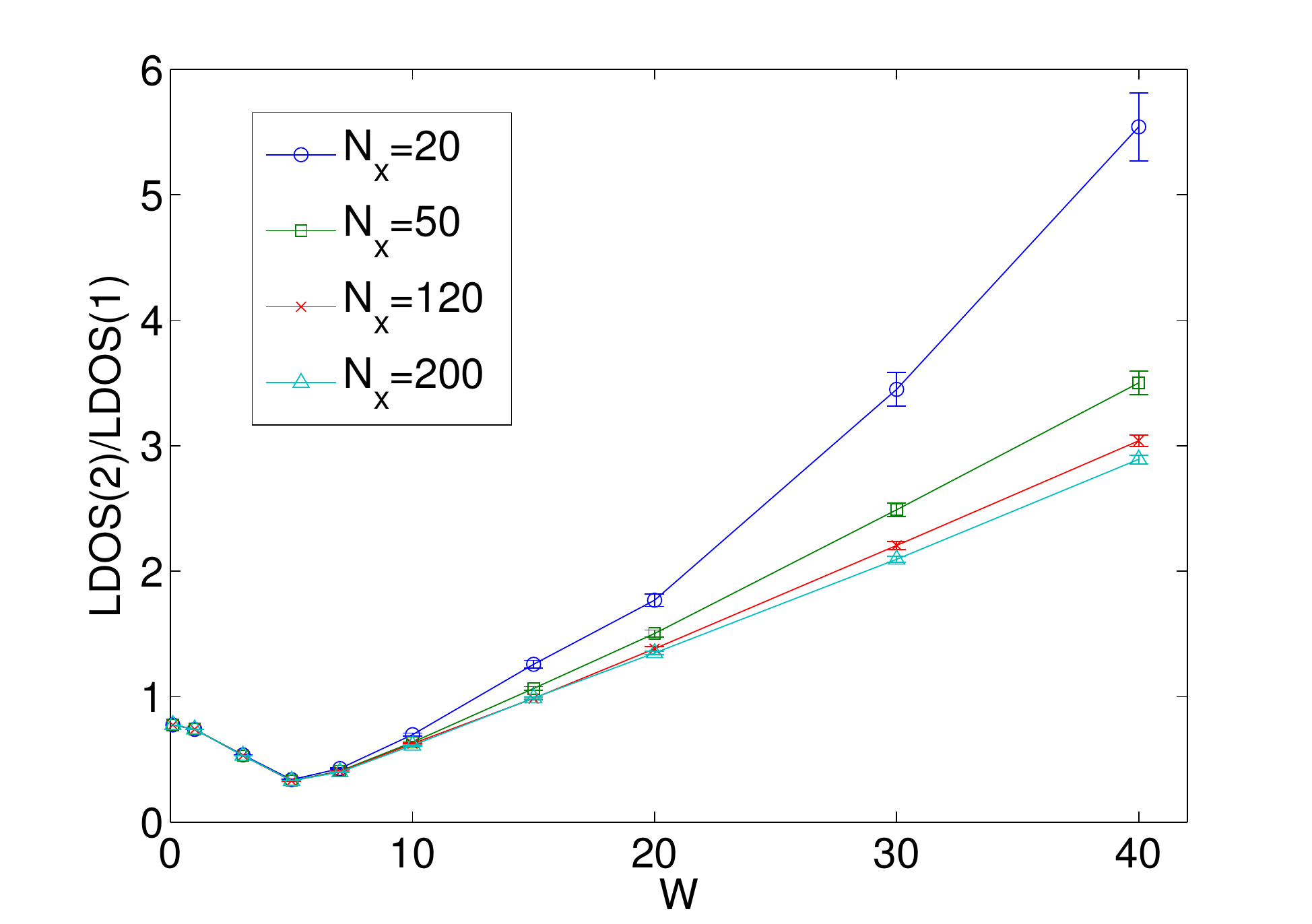}
\caption{The LDOS ratio of the first and second layer is plotted over the impurity strength $W$ for system size $N_x=20, 50, 120 , 200$. The edge state at zero energy is primarily populated in the first layer from weak to moderate impurity strength, $W<5$, and then the edge state moves to the second and following layers at strong impurity strength. The $N_x=200$ curve shows the behavior in thermodynamic limit as it becomes size-independent, while $N_x=20$ curve shows quantum-dot behavior for $W>5$, meaning the system size is not big enough to see bulk properties.}
\label{fig:ldosratio}
\end{figure}

In the strong impurity regime we must use a large enough system size to correctly see the size independent behavior of bulk electronic properties. Here, we distinguish the quantum-dot regime from the bulk regime by the dependence of physical observables on the system size. Figure \ref{fig:ldos_Nx} shows the histogram of the ratio of the local density of state in the first and the second wire for impurity strength $W=20$ with increasing system sizes $N_x=10,20,40,80$. The series of histograms shows size-dependence for $N_x \lesssim 40$, the histogram becomes Gaussian shape and size-independent for $N_x \gtrsim 40$. Therefore, if one wants to numerically obtain physical observables in the thermodynamic limit, it is important to use system size larger than $N_x=80$ for $W=20$ non-magnetic edge impurities.
The large-size requirement is less stringent for smaller impurity strengths, as evident from Fig. \ref{fig:ldosratio}. 


Figure \ref{fig:ldosratio} shows the ratio of the LDOS at zero energy with increasing impurity strength. This quantity tells us where the edge-state wavefunctions actually reside. A ratio below 1 indicates edge states rooted in the first layer. But ratios greater than 1 indicate edge states expelled to the the second layer, which are therefore increasingly less immune to the disorder. 

The edge states in weak and moderate disorder edge state are dominated by the first layer (also see Figure \ref{fig:vmap}). This comes hand in hand with a spread of the edge function Fourier transform: it has broad support away from $k_x=0$ due to impurity scattering. Once the impurity strength is comparable to or larger than the band width, the edge state is populated less in the impurity layer, and it moves to the second and following layers. All curves in Figure \ref{fig:ldosratio} show this behavior with a dip at $W=5$. While we believe that $N_x=200$ curve properly describes the system-size independent LDOS ratio in thermodynamic limit, $N_x=20$ curve is only good for $W<5$ and it begins to deviate from $N_x=200$ curve for strong impurity strength $W>5$.


\sub{Transfer matrix between single-layer wavefunctions}

Once the wavefunction of the first layer, $\psi_1$, is obtained, we next propagate it to the subsequent layers to obtain a full profile of the state. This can be done by using the matrices $M_n$ as in Eq.~\eqref{tran}. Let us write down the expression of the transfer matrix for the $(k_x,E)=(0,0)$ case first from layer m to layer n
\beq
T_{n\leftarrow m}&=& M_n M_m^{-1} \nonumber \\
&=& \left[ \lambda^{n-m} \ket{+}\bra{+}\,+\, \lambda^{m-n}\ket{- }\bra{-} \right] \frac{\sin (N+1-n)\theta}{\sin(N+1-m)\theta},\nonumber 
\eeq
where $\rho_{1,3} = \lambda e^{\pm i\theta}$ is used as before with $\lambda <1$, $N$ is the number of layers. We can see that the $\ket{+}$ component exponentially decays from the top surface towards the other end ($n>m$), while $\ket{-}$, if it is present, exponentially increases. Thus, it is apparent that $\ket{+}$ is a state localized to the top edge, while $\ket{-}$ is localized to the bottom edge. However, note that the transfer matrix contains an oscillating term dependent of the system size $N$, just as in the effective single-layer Hamiltonian, Eq.~\eqref{Heff}. It implies that if there is a $\ket{+}$ component in the wavefunction at exactly $(k_x, E)=(0,0)$, its oscillating part is dependent on the number of layers, which doesn't make sense in the physical picture where $N$ is much larger than the localization length of edge state. Therefore, we can say $\ket{+}$ component at $(k_x, E)=(0,0)$ must be vanishingly small as the system size is increased.

The transfer matrix from the first layer to the $n$\textsuperscript{th} layer for a general $(k_x, E)$ is expressed in the following way: 
\begin{align}\begin{split}
\bra{+}T_{n\leftarrow 1} \ket{+} &= \frac{1}{r_1-r_3} \left[ \rho_1^{n-1} r_1 - \rho_3^{n-1} r_3 \right] , \\
\bra{+}T_{n\leftarrow1} \ket{-} &= \frac{1}{r_1-r_3} \left[ \rho^{n-1}_1- \rho^{n-1}_3 \right] , \\
\bra{-}T_{n\leftarrow1} \ket{+} &= \frac{1}{1/r_3-1/r_1} \left[ \rho^{n-1}_1-\rho^{n-1}_3 \right] , \\
\bra{-}T_{n\leftarrow1} \ket{-} &= \frac{1}{1/r_3-1/r_1} \left[ \rho^{n-1}_1 /r_1 - \rho^{n-1}_3/ r_3 \right]. 
\end{split}\end{align}
with $r_i=\dirac{+}{\phi_i}/ \dirac{-}{\phi_i}$.

We apply this transfer matrix to the first layer wavefunction to obtain wavefunctions in the bulk layers. In Fig.~\ref{fig:vmap}, the disorder-averaged probability density in momentum and layer basis $P(k_x,n)=|\psi_n(k_x)|^2$ is plotted for shown impurity strength $W$. In the weak disorder regime, $W=0.1$, where impurity strength is much smaller than the energy gap, the probability density is concentrated near $k_x=0$ at zero energy. As the impurity strength increases the probability density gains width in momentum space and its weight is shifted to the second layer. While this trend is quite strong already with $W=10$, in strong impurity regime $W=40$---which is much larger than the bandwidth---the zero energy wavefunction is completely absent in the first layer, but occupies the subsequent layers in a narrow range of momentum space.
This indicates that the wavefunction has been pushed to the next layer and behaves as if the system is clean.

This behavior of the local density of states is shown for non-magnetic edge impurities, which cannot affect the transport properties of helical edge states in 2D topological insulators. Therefore, the modification of LDOS should be discussed separately from the change of transport nature, at least in 2D. For a strip geometry, the transport is studied for both non-magnetic and magnetic edge impurities in the following sections.

The widths of probability density $P(k_x,n)=|\psi_n(k_x)|^2$ in the first and second layer at zero energy are plotted in Fig.~\ref{fig:width}. The width in momentum space is indicative of how disordered the edge state is due to impurities.
In the weak and strong disorder limit the wavefunction behaves like a clean system in the LDOS shape as shown in Fig.~\ref{fig:vmap}.
With strong disorder, the width of $P(k_x,n=1)$ saturates near $0.3$, although it carries little weight in that limit: $P(k_x,n=1) \ll 1$.
Meanwhile, the width of $P(k_x,n=2)$ increases and decreases again as the impurity strength is varied, peaking at around the bandwidth of the system ($W\sim8$).


\subsection{Magnetic edge impurities}

\begin{figure}[t]
\begin{center}
\includegraphics[width=85mm]{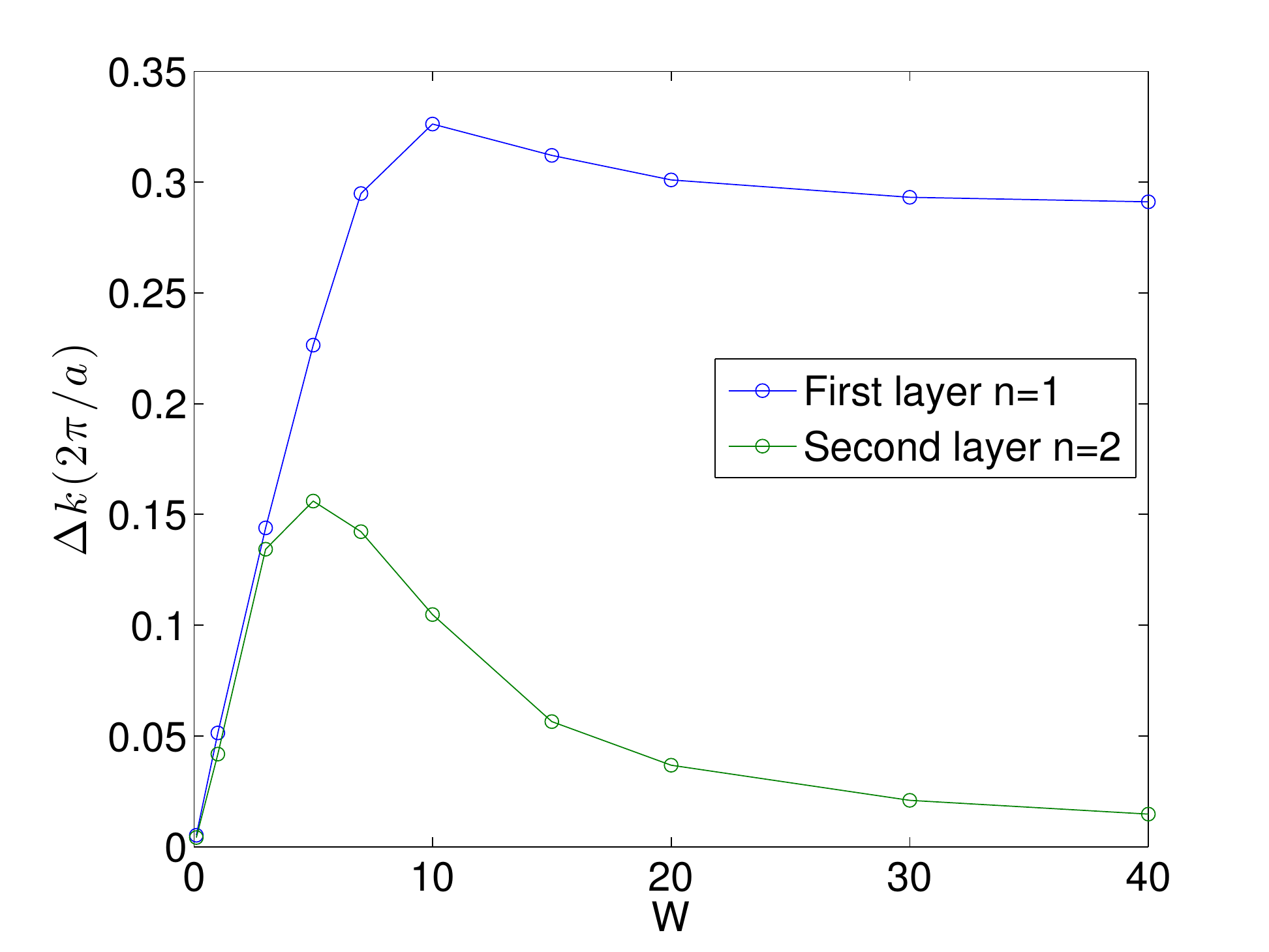}
\caption{
LDOS momentum-space width in the first and the second layers with {\it magnetic} impurities, at system size $N_x=120$. The overall behavior is the same with the case of non-magnetic edge impurity case (Fig.~\ref{fig:width}), but much larger systems are required to see the size-independent physical properties in bulk regime. }
\label{fig:width_mag}
\end{center}
\end{figure}

The same calculation was repeated for a system with magnetic edge impurities. We simply needed to extend the Hamiltonian to have two spin-sections and introduce random magnetic impurities, $V(x_i) = \vec{V_i} \cdot \vec{s}$, where three component random variable $\vec{V_i} = (V_{i}^{x},V_{i}^{y},V_{i}^{z})$. We found that to simulate the bulk regime for $W=20$ the system size needs to be at least $N_x=400$ as opposed to $N_x=120$ for the non-magnetic edge impurity case. In other words, the lower bound of the system size to see the thermodynamic properties is much larger and it becomes computationally challenging even for 2D system. 
Figure~\ref{fig:width_mag} shows the width of the LDOS distribution $P(k_x,n)$ for $n=1,2$.
The data is qualitatively similar to the non-magnetic case, which demonstrates the universality of the result.

\section{Transport behavior}\label{sec:tran_beha}

The local density of states discussed in the last section can be probed by angle-resolved photoemission spectroscopy.\cite{wray2011}
The transport along the edge states, however, is provides additional information independent of the local density of states. For instance, in the case of non-magnetic impurities, edge modes can not backscatter, and their conductance remains quantized at the value of the clean system, despite the local density of states associated with it changing its support between the layers. To clarify the transport nature of the system with magnetic and non-magnetic impurities along the edge, in this section we study conductance of the systems using Landauer-B{\"u}ttiker method. 


Imagine a system where two semi-infinite leads are connected to a disordered region at the ends $x=1$ and $x=N_x$. Landauer and B{\"u"ttiker \cite{land1957,land1970,butt1988} related the conductance with the transmission coefficient through the disordered region: $g = \frac{e^2}{h} T_{N1} $, where $g$ is conductance, $T_{N1}$ is a transmission coefficient through the disordered region from site 1 to site $N$. Using linear response formalism, Fisher and Lee \cite{fish1981} expressed the transmission coefficient in terms of Green's functions: 
\beq g = \frac{e^2}{h}\operatorname{Tr}\left[ \Gamma_L G_{N1} \Gamma_R G_{N1}^\dg \right] \eeq
where $\Gamma_L= i (\Sigma_L - \Sigma_L^\dg)$, $G_{N1}$ is a Green's function from site 1 to $N$ renormalized by the presence of the leads, and $\Sigma_L$ is a self-energy of the semi-infinite left lead. Each term in this formula can be computed recursively, such that the conductance of a long system can be obtained with a reasonable computation effort. A good review of the detailed calculation can be found in Ref.~\onlinecite{meta2005}.

\sub{Non-magnetic impurities case}\label{sec:non_impu}

Consider the 2D topological insulator system introduced earlier with non-magnetic impurities along the top edge. Because the Hamiltonian is diagonal in spin basis without magnetic impurities, we can consider transport in just one spin sector. When the chemical potential is in the energy gap, neither backscattering nor scattering into the bulk is possible. Therefore, the conductance must remain quantized even in the presence of edge impurities. This is indeed what our calculation shows. Indeed, when the disorder is non magnetic, the transport behavior does not reflect the development of the LDOS.

\sub{Magnetic impurities case}\label{sec:magn_impu}

The two opposite-spin, counter-propagating, chiral edge modes couple as soon as magnetic edge impurities are introduced. As a result, transport through the disordered edge is suppressed, while the transport through the clean edge remains unaffected. Therefore, we expect the total conductance to rapidly approaches $e^2/h$ when introducing and increasing magnetic edge disorder.

However, in the strong-impurity quantum-dot limit, $W \gg N_x$, we found that the conductance recovers its clean system value of $2e^2/h$. In this regime the impurities are strongly bound to electrons at energies far away from the Fermi energy, and they play negligible roles in the transport at the Fermi energy. Put another way, strong disorder pushes the edge modes to the next layer where they effectively become weak scatters. This behavior is clear in Fig.~\ref{fig:spinful3_W}, which shows the conductance vs.\ the disorder strength for different system sizes. We can see that in the intermediate range of impurity strength, the conduction through the disordered top edge is significantly suppressed due to magnetic impurities, while the conductance recovers up to $2e^2/h$ value in the strong impurity limit. 
We note that in the thermodynamic limit $N_x \rightarrow \infty$, the conductance is always $e^2/h$ for any disorder strength, illustrated in Fig.~\ref{fig:spinful3_dav_W}. Our calculation reflects the non-monotonic dependence of the localization length of the scattered edge mode on disorder strength, which is consistent with the expulsion of the LDOS from the disordered first layer.


\begin{figure}[h]
\includegraphics[width=85mm]{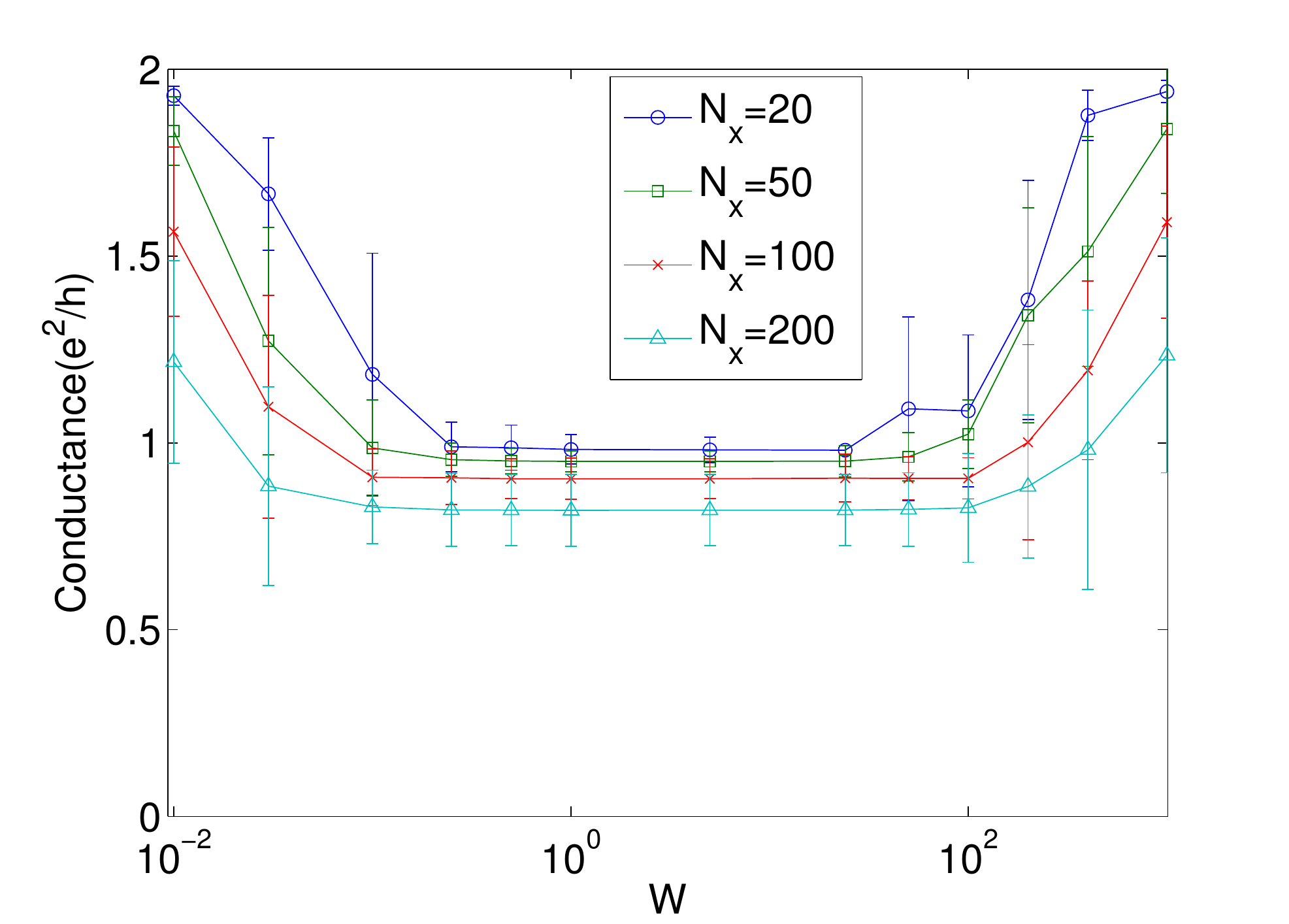}
\caption{Edge channel conductance vs. impurity strength when magnetic edge impurities are present. Three system sizes $N_x=20, 50, 100, 200$ are used. The edge channel conductance drops to $e^2/h$ initially. The conductance recovers to $2e^2/h$ as the system enters quantum-dot regime at strong magnetic impurity strength. This reflects the nonmonotonic behavior of the localization length of the edge mode as disorder increases. }
\label{fig:spinful3_W}
\end{figure}

\begin{figure}[h]
\includegraphics[width=85mm]{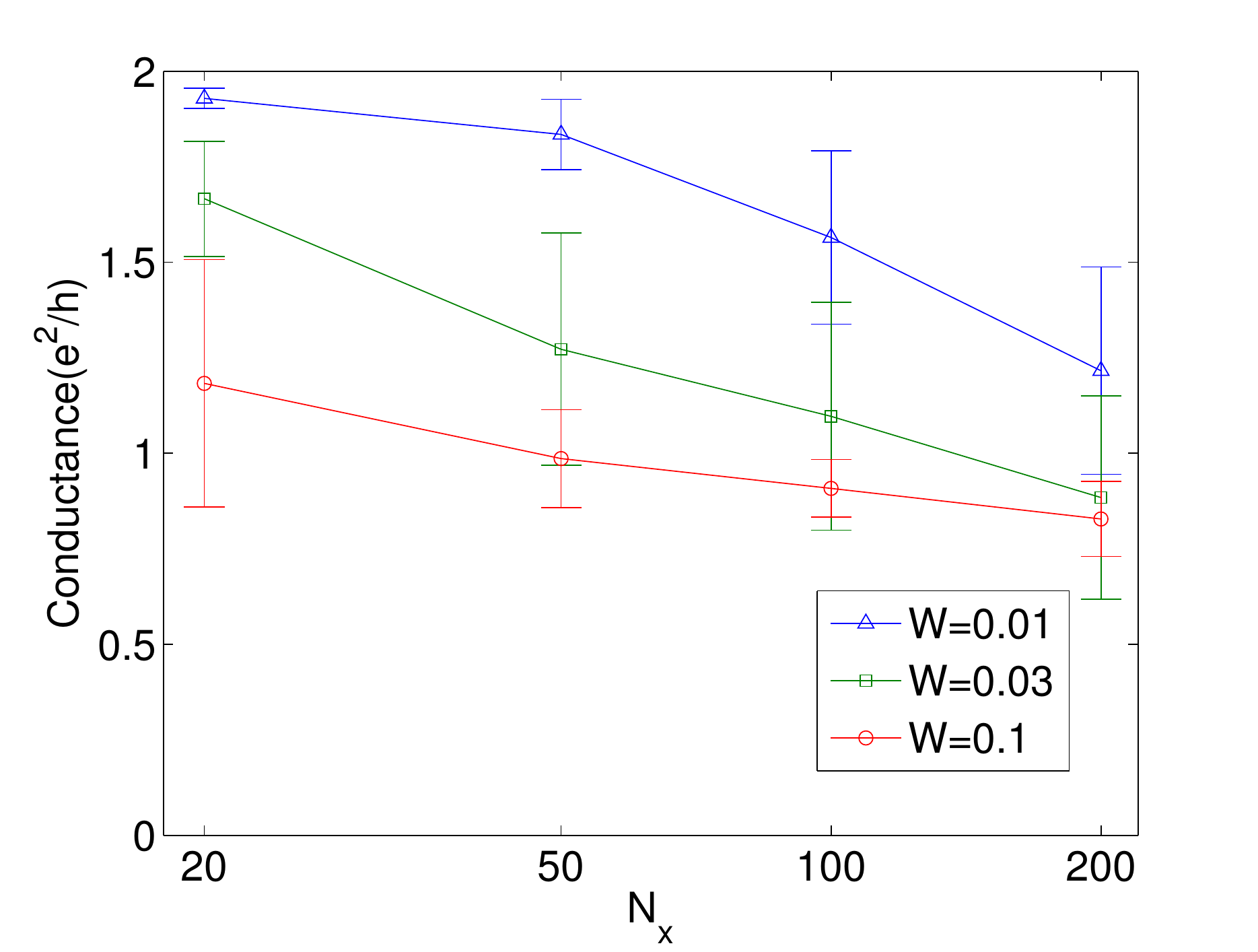}
\caption{Disorder averaged conductance with increasing system size $N_x$ for three impurity strength $W$. The conductance exponentially decreases with system size as anticipated in the magnetic edge impurity case. }
\label{fig:spinful3_dav_W}
\end{figure}

\section{Summary and Conclusion}\label{sec:conc_and}

The effects of surface disorder on systems with surface states are at the focus of our work. The interest in this problem rose after an exact-diagonalization analysis showed that the response of a 3D topological insulator to surface disorder is non-monotonic: First the surface states become diffusive, and their conductance is suppressed. At a finite disorder strength, however, the surface states mean free path recovers, and they reconstitute at the disorder-free second layer of the TI \cite{schu2012}.

In this manuscript we developed a formalism that reduces solving a bulk $D$-dimensional Hamiltonian to a surface-only diagonalization problem. This, in principle, enables an exact-diagonalization analysis of much bigger bulk systems than previously possible. Roughly speaking, our method constitutes a systematic integrating out and elimination of the bulk degrees of freedom layer by layer. We show how to carry this procedure out with a technique reminiscent of transfer-matrix methods. From the resulting surface-only diagonalization problem we are able to reconstruct the wavefunctions of the bulk layers and study the effect of disorder on them. Our method is generally valid for any system composed of coupled layers but becomes particularly useful for topological systems whose surface states are fundamentally dependent on the existence of the gapped bulk. 

We used our method to study 2D topological insulators with edge impurities. As the strength of edge impurities increases, we found that edge states near zero energy become more concentrated up to the moderate impurity strength (i.e., comparable to the bandwidth), and then the edge states are gradually pushed to the second layer. The ratio of the local density of states in the first two layers in Figure \ref{fig:ldosratio} shows this behavior for different system sizes. Furthermore, the width of the edge state in momentum space in the second layer reaches a maximum at moderate impurity strength and then narrows at stronger impurity strength. Despite the strong disorder, the edge states have momentum restored to being a good quantum number. This non-monotonic response to the surface disorder is equally true for magnetic and non-magnetic disorder (see Figures \ref{fig:width} and \ref{fig:width_mag}).

The transport properties, however, show a sharp contrast between the magnetic and non-magnetic cases: non-magnetic edge impurities does not affect the edge transport properties, while magnetic edge impurities immediately induce a finite localization length due to backscattering, and initially suppress the edge conductance (see Figure \ref{fig:spinful3_W}). In our numerical simulations that include magnetic impurities, however, we found that the conductance then recovers to the clean-system values when the impurity strength exceeds the system size. 

This finding does not imply that the edge states completely decouple at large but finite disorder. Rather, it is an indication that the localization length of the spin-orbit locked modes exhibits a non-monotonic localization behavior as a function of disorder. In our simulations, perfect conductance will be recovered when the localization length exceeds the system width. This is precisely the regime we nicknamed the quantum-dot regime. The localization length inferred from transport calculations must be proportional to the inverse of the average width of the edge LDOS in momentum space, for momenta parallel to the edge. Therefore, our transport results are yet another manifestation of the reconstitution of the edge mode at the second layer, which is disorder free.
In the limit of infinitely strong disorder, we expect the localization length of the reconstituted edge state to diverge and translational invariance is regained.

In future work, we intend to apply our method to the 3D topological insulator with surface disorder. As we emphasize throughout, one must use large enough systems in order to explore the bulk properties in thermodynamic limit (see Figure \ref{fig:ldos_Nx} and \ref{fig:ldosratio}). This makes 3D topological systems with moderate to strong surface impurities harder to study using the ED as the cost of calculation increases. Our analytic approach will then be very useful, since it reduces the computational cost dramatically.


\section*{Acknowledgments}
It is a pleasure to acknowledge support from the Sherman-Fairchild Foundation (RM), the Packard Foundation, the Walter Burke Institue of Theoretical Physics, as well as the Institute of Quantum Information and Matter, an NSF frontier center, with the support of the Gordon and Betty Moore Foundation (KWK, GR). In addition, support from NSERC and CIFAR is gratefully acknowledged (MF).

\clearpage
\appendix
\section*{Appendix: derivation of the effective Hamiltonian and the transfer matrix}\label{sec:appe}

The construction of the object $M$ in Sec.~\ref{sec:gene_fram} is a central element for further computation in any examples. In this appendix, we show how the object $M$ in Sec.~\ref{sec:appl_to} was constructed for given momentum $k_x$ and energy $E$. We work in the momentum space since the clean Hamiltonian is diagonalized in momentum space. But, as the layer degree of freedom is integrated out, the effective Hamiltonian is dependent of energy nontrivially as can be seen in the recursive Eq.~\eqref{recu_back}. Therefore, the dispersion relation cannot be immediately deduced from the single layer effective Hamiltonian. Rather, we compute the quantity $\det [E-H_1^\textrm{eff}-V_\textrm{imp} ] $ in momentum space as varying energy $E$ such that the determinant is zero. In this way we find eigenenergies and then eigenstates of a surface disordered system.

The object $M$ satisfies the layered Schrodinger equation Eq.~\eqref{scho_M}. Thus, the construction is convenient in terms of the wavefunctions of the Schr\"odinger equation, Eq.~\eqref{scho}. The most general expression will be
\beq
M_{l+N+1} = \sum_{i,j=1,2,3,4} c_{ij} \rho_i^{l} \ket{\phi_i} \bra{\phi_j},
\eeq
with the vanishing boundary condition at the last layer $M_{N+1} = 0$. There are different ways to build $M$ and we introduce one way for on-shell $E=A \sin k_x$ and off-shell $E \neq A \sin k_x$ states.

\sub{On-shell states, $E = A\sin k_x$}

Considering only one spin sector of the Hamiltonian as in Sec.~\ref{subsection:half_syst}, the eigenvalues and eigenvectors of the transfer matrix over the layer are straightforwardly obtained from the Schr\"odinger equation with the replacement $\psi_{n-1} = \frac{1}{\rho} \psi_n$ and $\psi_{n+1} = \rho \psi_n$, 
\beq
B\frac{1}{\rho}\psi_{n} + [H_{0}(k_x)-E]\psi_{n} + B^\dagger \rho \psi_{n}=0. \label{scho2}
\eeq
Solving a 2x2 matrix equation, we obtain four eigenvalues $\rho_{i=1,2,3,4}$ with eigenvectors $\phi_{i=1,2,3,4}$. For on-shell states, we immediately know that two eigenvectors are $\ket{+}$,
\beq \phi_1 = \bpm \alpha_1 \\ \beta_1 \epm, \phi_2 = \bpm \alpha_2 \\ \beta_2 \epm, \phi_3 = \frac{1}{\sqrt{2}}\bpm 1 \\ 1 \epm, \phi_4 = \frac{1}{\sqrt{2}}\bpm 1 \\ 1 \epm, \nonumber \eeq
with eigenvalues $| \rho_{i=3,4}| <1$ so that $\phi_{j=3,4}$ are physically localized states at the top layer $n=1$. This is to satisfy the vanishing boundary condition of the clean system with two edge-localized wavefunctions. To construct the object $M$, we need not only the vanishing boundary condition, but also $M$ must be invertible. To do that, consider the decomposition of $\phi_{i=1,2}$ into $\ket{+}$ and $\ket{-}$: 
\beq \phi_{j=1,2} = \bpm \alpha_j \\ \beta_j \epm = \frac{A_{j}}{\sqrt{2}} \bpm 1 \\ 1 \epm + \frac{B_{j}}{\sqrt{2}}\bpm 1 \\ -1 \epm \label{schm} \label{deco} \eeq
From this set of eigenvectors, we can construct two copies of $\ket{-}$'s behaving differently over the layers. Specifically, 
\beq
\ket{-_{(j,l)}} = \frac{1}{B_j}\phi_{j} - \frac{A_j}{B_j} \phi_{l}
\eeq
where $j=1,2$ and $l=3,4$. Then, we can think the following construction of the object $M$: 
\beq
M_{N+1} = \left[ \ket{-_{(1,3)}} - \ket{-_{(2,4)}} \right] \bra{-} + \left[ \phi_3 -\phi_4 \right] \bra{+} \nonumber
\eeq
which is zero. For $n \in [0,N]$, $M_n$ is of course non-zero as eigenvalues of the eigenvectors are different, and $M_n$ is generally invertible. For $n_{th}$ layer, 
\begin{subequations}\begin{align}
\bra{+} M_n \ket{+} &=& \rho_3^{n-N-1} - \rho_4^{n-N-1}, \\
\bra{+} M_n \ket{-} &=& \frac{A_1}{B_1}\left[ \rho_1^{n-N-1} - \rho_3^{n-N-1} \right] \notag\\
&&+ \frac{A_2}{B_2}\left[ \rho_4^{n-N-1} - \rho_2^{n-N-1} \right] \\
\bra{-} M_n \ket{+} &=& 0 \\
\bra{-} M_n \ket{-} &=& \rho_1^{n-N-1} - \rho_2^{n-N-1}
\end{align}\end{subequations}
From this, the construction of the transfer matrix and the effective single layer Hamiltonian is following:
$T_{n \leftarrow m} = M_{n} M_m ^{-1}$: 
\begin{subequations}\begin{align}
&\bra{+} T_{n \leftarrow m} \ket{+} = \frac{\rho_3^{n-N-1}-\rho_4^{n-N-1}}{\rho_3^{m-N-1}-\rho_4^{m-N-1}}\\
&\bra{+} T_{n \leftarrow m} \ket{-} = \frac{\frac{A_1}{B_1}\rho_1^{n-N-1}-\frac{A_2}{B_2}\rho_2^{n-N-1}}{\rho_1^{m-N-1}-\rho_2^{m-N-1}} \nonumber \\
&+ \frac{(\rho_3^{n-N-1}-\rho_4^{n-N-1})\left( \frac{A_2}{B_2}\rho_2^{m-N-1}- \frac{A_1}{B_1}\rho_1^{m-N-1} \right) }{(\rho_3^{m-N-1}-\rho_4^{m-N-1})(\rho_1^{m-N-1}-\rho_2^{m-N-1})}\\
&\bra{-} T_{n \leftarrow m} \ket{+}= 0 \\
&\bra{-} T_{n \leftarrow m} \ket{-} = \frac{\rho_1^{n-N-1}-\rho_2^{n-N-1}}{\rho_1^{m-N-1}-\rho_2^{m-N-1}}
\end{align}\end{subequations}
Expressing $B = (b+A/2) \ket{+}\bra{-} + (b-A/2) \ket{-}\bra{+}$, the effective single layer Hamiltonian is:
\begin{subequations}\begin{align}
\bra{+} E-H_1^\textrm{eff}\ket{+} &=& (b-A/2)\bra{-} T_{0 \leftarrow 1} \ket{+} \label{H1} \\
\bra{+} E-H_1^\textrm{eff}\ket{-} &=& (b-A/2)\bra{-} T_{0 \leftarrow 1} \ket{-} \\
\bra{-} E- H_1^\textrm{eff}\ket{+} &=& (b+A/2)\bra{+} T_{0 \leftarrow 1} \ket{+} \\
\bra{-} E- H_1^\textrm{eff}\ket{-} &=& (b+A/2)\bra{+} T_{0 \leftarrow 1} \ket{+} \label{H4}
\end{align}\end{subequations}
One can see that the effective Hamiltonian is still system-size dependent as $H_1^\textrm{eff}(k_x=0,E=0)$ case computed in the main manuscript. Plus, $E-H_1^\textrm{eff}$ contains no zero eigenvalues, implying that any finite system cannot have the energy dispersion $E=A \sin k_x$ due to finite-size effects. Therefore, the expression of the Hamiltonian for on-shell states is not useful for actual computation. Rather, we need the Hamiltonian expression of off-shell states for finites size system with surface impurities, for which we find an analytic expression of the effective single layer Hamiltonian and let $N \rightarrow \infty$ for the semi-infinite limit.

\begin{widetext}
\sub{Off-shell states, $E \neq A\sin k_x$}

For a system with surface impurities, eigenstates are not described by the clean system dispersion $E=A\sin k_x$, rather, the state has a mix of different momentum components in each given energy. More concretely, if the size of the system along the periodic boundary condition is $N_x$, the exact edge state dispersion of the clean system discussed in Sec.~\ref{sec:appl_to} is $E= A \sin (2\pi l/N_x)$ with integer $l$. Only at those discrete set of energies, we have two eigenvectors parallel to $\ket{+}$ and the discussion in the previous section applies. Except those on-shell points, we have the following general set of eigenvectors,
\begin{align}
	\phi_1 = \bpm \alpha_1 \\ \beta_1 \epm,\quad
	\phi_2 = \bpm \alpha_2 \\ \beta_2 \epm,\quad
	\phi_3 = \bpm \alpha_3 \\ \beta_3 \epm,\quad
	\phi_4 = \bpm \alpha_4 \\ \beta_4 \epm
\end{align}
with eigenvalues $\rho_{j=1,2,3,4}$. Without loss of generality, let us say $|\rho_{j=2,4}|>1$. Each eigenvectors can be written as the sum of $\ket{+}$ and $\ket{-}$ like Eq.~\eqref{deco}. Then, we similarly construct two pairs of $\ket{+}$ and $\ket{-}$ by the superposition of $\phi_{j=1,2,3,4}$. 
\begin{align}
	\ket{+_{(j,l)}} &= \frac{1}{\frac{A_j}{B_j}-\frac{A_l}{B_l}}\left( \frac{1}{B_j}\phi_j - \frac{1}{B_l} \phi_l \right),
&	\ket{-_{(j,l)}} &= \frac{1}{\frac{B_j}{A_j}-\frac{B_l}{A_l}}\left( \frac{1}{A_j}\phi_j - \frac{1}{A_l} \phi_l \right).
\end{align}
Next, we can construct the object $M_n$ satisfying the vanishing boundary condition at $n=N+1$. 
\begin{align}
	M_{N+1} = \big( \ket{-_{(1,2)}} - \ket{-_{(3,4)}} \big) \bra{-} + \big( \ket{+_{(1,2)}} - \ket{+_{(3,4)}} \big) \bra{+}
\end{align}
where we intentionally split $\phi_2$ and $\phi_4$ in each term so that no terms vanish in $N \rightarrow \infty$ limit. Explicitly, the components are
\begin{subequations}\begin{align}
	\bra{+}M_{l+N+1} \ket{+} &= \frac{1}{\frac{A_1}{B_1}- \frac{A_2}{B_2}}\left[ \frac{A_1}{B_1}\rho_1^l - \frac{A_2}{B_2}\rho_2^l \right]
		- \frac{1}{\frac{A_3}{B_3}- \frac{A_4}{B_4}}\left[ \frac{A_3}{B_3}\rho_3^l - \frac{A_4}{B_4}\rho_4^l \right] ,
\\	\bra{-}M_{l+N+1} \ket{+} &= \frac{1}{\frac{A_1}{B_1}- \frac{A_2}{B_2}}\left[ \rho_1^l - \rho_2^l \right]
		- \frac{1}{\frac{A_3}{B_3}- \frac{A_4}{B_4}}\left[ \rho_3^l - \rho_4^l \right] ,
\\	\bra{+}M_{l+N+1} \ket{-} &= \frac{1}{\frac{B_1}{A_1}- \frac{B_2}{A_2}}\left[ \rho_1^l - \rho_2^l \right]
		- \frac{1}{\frac{B_3}{A_3}- \frac{B_4}{A_4}}\left[ \rho_3^l - \rho_4^l \right] ,
\\	\bra{-}M_{l+N+1} \ket{-} &= \frac{1}{\frac{B_1}{A_1}- \frac{B_2}{A_2}}\left[\frac{B_1}{A_1} \rho_1^l - \frac{B_2}{A_2}\rho_2^l \right]
		- \frac{1}{\frac{B_3}{A_3} - \frac{B_4}{A_4}}\left[\frac{B_3}{A_3} \rho_3^l - \frac{B_4}{A_4}\rho_4^l \right] .
\end{align}\end{subequations}
We are interested in the behavior of edge states near the top ($n=1$). Thus, in the limit of $N \rightarrow \infty$, $l=n-N-1 \rightarrow -\infty $ and the terms with $\left(\frac{\rho_{j=1,3}}{\rho_{j=2,4}}\right)^{l}$ dominates.
\begin{align}\begin{alignedat}{3}
		\bra{+} T_{n \leftarrow m} \ket{+} &= \frac{\frac{A_1}{B_1}\rho_1^{n-m}-\frac{A_3}{B_3}\rho_3^{n-m}}{\frac{A_1}{B_1}- \frac{A_3}{B_3}} ,
	&\qquad	\bra{+} T_{n \leftarrow m} \ket{-} &= \frac{\rho_1^{n-m}-\rho_3^{n-m}}{\frac{A_1}{B_1}- \frac{A_3}{B_3}} ,
	\\	\bra{-} T_{n \leftarrow m} \ket{+} &= \frac{\rho_3^{n-m}-\rho_1^{n-m}}{\frac{B_1}{A_1}- \frac{B_3}{A_3}} ,
	&\qquad	\bra{-} T_{n \leftarrow m} \ket{-} &= \frac{\frac{B_3}{A_3}\rho_3^{n-m}-\frac{B_1}{A_1}\rho_1^{n-m}}{\frac{B_1}{A_1}- \frac{B_3}{A_3}} .
\end{alignedat}\end{align}
And the expression for the Hamiltonian is following Eq.~\eqref{H1}--\eqref{H4}. 

\end{widetext}

\bibliography{kun_biblio}

\end{document}